\documentclass[10pt]{amsart}

\usepackage{geometry} 
\usepackage{graphicx,url, hyperref}
\usepackage{amssymb,amsmath, amsthm}
\usepackage{epstopdf}
\usepackage{subcaption}
\usepackage[utf8]{inputenc} 

\usepackage{multirow} 
\usepackage{array}

\numberwithin{equation}{section}

\newcommand{\E}{{\mathbf e}}

\usepackage{dsfont}
\usepackage{enumerate}
\theoremstyle{remark}
\newtheorem{remark}{Remark}[section]

\usepackage{xcolor}
\usepackage[percent]{overpic}

\begin{document}

\title{Anomalous random flights and time-fractional run-and-tumble equations}

\author{Luca Angelani}
\address{ Institute for Complex Systems (ISC), CNR, and 
Department of Physcs,  ``Sapienza" University of Rome, P.le A. Moro 2, 00185 Rome, Italy}
\email{luca.angelani@cnr.it}


\author{Alessandro De Gregorio}
\address{Department of Statistical Sciences, ``Sapienza" University of Rome, P.le Aldo Moro, 5 - 00185, Rome, Italy}
\email{alessandro.degregorio@uniroma1.it}

\author{Roberto Garra}

\address{Section of Mathematics, International Telematic University Uninettuno, Corso Vittorio Emanuele II, 39, 00186 Roma, Italy}
\email{roberto.garra@uninettunouniversity.net}

\author{Francesco Iafrate}
\address{Department of Basic and Applied Sciences for Engineering, ``Sapienza" University of Rome,
	Via Antonio Scarpa, 14 - 00161, Rome, Italy}
\email{francesco.iafrate@uniroma1.it}



\maketitle

\begin{abstract}
Random flights (also called run-and-tumble walks or transport processes) represent finite velocity random motions changing direction at any Poissonian time.  These models in $d$-dimension, can be studied giving a general formulation of the problem valid at any spatial dimension.
The aim of this paper is to extend this general analysis to time-fractional processes arising from a non-local generalization of the kinetic equations.
The probabilistic interpretation of the solution of the time-fractional equations leads to a time-changed version of the original transport processes. The obtained results provide a clear picture of the role played by the time-fractional derivatives in this kind of random motions. They display an anomalous behavior and are useful to describe several complex systems arising in statistical physics and biology. 
In particular, we focus on the one-dimensional random flight, called telegraph process, studying the time-fractional version of the classical telegraph equation and providing a suitable interpretation of its  stochastic solutions.
\end{abstract}


\section{Introduction}

Random flights (or transport processes) describing the finite velocity random motion of a particle in a $d$-dimensional space have been object of many studies in the probabilistic and physical literature. 
There are many different models related to these random motions of a particle in $\mathbb R^d$. The first formulation probably dates back to Pearson which considered a random walk with fixed and constant steps 
\cite{Pearson}. Many papers appeared in literature analyzed isotropic random motions with finite velocity choosing new direction uniformly on a sphere at each Poisson jumping time; see, for example,  \cite{monin1956statistical, stadje1987exact, stadje1989exact, kolesnik2005planar, orsingher2007random}. Furthermore, the kinetic equations represent a useful tool to describe transport processes. Some generalizations of the latter models have been proposed by assuming non-uniform scattering mode and/or  time steps with more general probability distributions  (see, e.g., \cite{le2010pearson, de2012flying, de2012random, pogorui2012random}). It is particularly interesting the one-dimensional model, also called {\it telegraph process}, introduced in \cite{goldstein1951diffusion,kac1974stochastic}; 
in this case, at Poissonian random time instants, 
the particle reverses its direction of motion 
and then admits only two possible directions. Furthermore, the probability law of the position reached from the particle at time $t>0$ is solution of the  telegraph equation (see \cite{weiss2002some} and references therein). A complex version of the telegraph process has been studied in \cite{de2021telegraph}.

On the other hand, in the physical literature, run-and-tumble motions are particular random flights widely used for the study of active particles, for example to describe the dynamics of motile bacteria, such as {\it E.coli}
\cite{Ecoli_Berg,RWinBio_Berg,PhysRevE.48.2553,RevModPhys.88.045006,martens2012probability,PhysRevLett.100.218103,Cates_2012,angelani2013averaged,PhysRevE.48.939}.
The motion of run-and-tumble particles alternates stochastic time periods during
which the particle moves along a randomly chosen
direction. For these reasons, it can be considered as a persistent time random walk \cite{weiss2002some}. Also in this case there is a growing literature in which run-and-tumble models are applied in 
a variety of different contexts and physical situations, such as, for example, 
to investigate geometrical confinement and escape problems  
\cite{angelani2023one,Angelani2024EPJE,Guéneau_2024,Bressloff_2022,Bressloff_2023},
irreversible trapping  \cite{Angelani_2023},
resetting processes \cite{Evans_2018,PhysRevE.106.044127},
entropy production \cite{Garcia-Millan_2021,PhysRevE.105.034113,PhysRevE.108.014139}
or analyze experimental scattering functions of bacterial suspensions 
\cite{PhysRevLett.132.038302,PhysRevE.109.014612}
(just to mention a few very recent works on selected topics).

Furthermore, several complex systems exhibit nonlinear  mean-squared displacement over time, long-range correlations, nonexponential relaxation, heavy-tailed and skewed marginal distributions, lack of scale invariance, trapping effects (see, e.g., \cite{scher1975anomalous}). Therefore, such phenomena follow an ``anomalous'' dynamics and cannot be described by means of classical diffusion models. Fractional kinetic equations represent useful tools for the description of transport dynamics in complex systems, which are governed by anomalous diffusion (see, e.g., \cite{metzler2000random}). 

In the recent paper \cite{sevilla2023anomalous}, the authors have studied the time-fractional generalization of the kinetic equation in order to show the utility of fractional models to study anomalous transport problems of active particles. This fractional generalization of the run-and-tumble process is interesting to describe the transition from super- to sub-diffusive anomalous behaviours. Moreover, the fractional kinetic equation is directly related to the time-fractional telegraph-type equation that has been object of many mathematical studies in the recent literature (we refer, e.g., to \cite{orsingher2004time,d2014time,masoliver2021telegraphic,PhysRevE.101.012137} and the references therein).  
Anomalous phenomena have also been studied by means of generalized telegraph equations defined as integrodifferential equations with memory kernels responsible for the time smearing of the first and second time derivatives, respectively (see \cite{PhysRevE.102.022128}, \cite{PhysRevE.104.024113} and \cite{gorska2023subordination}).

Inspired by this model, in this paper, we provide a new and clear stochastic interpretation of the anomalous random flights governed by the fractional kinetic equation where the classical time derivative is replaced with fractional Caputo derivative; i.e. let $n\in{\mathbb N}^+,$ for a suitable function $f$ the fractional Caputo derivative is defined as follows
\begin{align} \label{de:fd}
\partial_t^\nu f(t)= 
\begin{cases}
    \frac{1}{\Gamma(n-\nu)}\int_0^t (t-\tau)^{n-1-\nu} \ \partial_\tau^n f(\tau)
    d\tau,& n-1<\nu<n,\\
  \partial_t^n f(t), & \nu=n ,  
\end{cases}
\end{align}
where $\partial_t^n f(t)$ denotes the ordinary time-derivative of order $n$
and $\Gamma(z)$ is the Euler gamma function.
If we consider $\nu\in(0,1)$ ($n=1$), the Laplace transform $\mathcal L$ of \eqref{de:fd} becomes
\begin{equation}\label{ltfd}
    {\mathcal L} [\partial_t^{\nu} f(t) ] (s)=s^\nu  {\mathcal L} [f(t) ] (s) 
    - s^{\nu-1} f(0)
\end{equation}
(the reader can consult the fractional calculus monograph \cite{kilbas2006theory}).

In the general $d$-dimensional case, we obtain a random flight time-changed with the inverse of stable subordinators (i.e. the first hitting time of an increasing and non-negative Lévy processes with Laplace exponent given by $\psi(u)=u^\nu,\nu\in(0,1)$). Indeed, we prove that the formulation of the fractional problem can be reduced to the general theory of time-changed random processes. We highlight that the transport process obtained from fractional kinetic equation is not still with finite velocity and has sample paths  trapped in some time intervals.  Furthermore, the particle shows nonlinear diffusion behavior over time.
Then, we consider in more detail the one dimensional case that is the more interesting and studied in the literature. 
First of all, we prove the relation between the fractional telegraph-type  equation and the fractional kinetic equation. Then, we obtain the stochastic solution of the fractional telegraph process that coincides with the time-changed telegraph process and generalizes the result obtained in the standard framework (see \cite{kac1974stochastic}). On this topic the reader can also consult the paper \cite{li2019fractional}, where the authors provide d’Alembert’s formulas for abstract fractional telegraph equations.

The paper is organized as follows. Section \ref{overview} contains an overview on the run-and-tumble motions in arbitrary $d$-dimension. In this section we show that, starting from the general kinetic equation, we can recover many interesting explicit non-trivial results present in the literature.
In Section \ref{fkequation}, we introduce a time-fractional linear Boltzmann equation; the main idea is to replace the classical derivative with the Caputo derivative and introduce the related random motions. By resorting to the general theory of non-local operators and time-changed random processes (briefly recalled in Appendix), in Section \ref{atprocesses}, we discuss the  interpretation of $d$-dimensional anomalous isotropic transport processes as time-changed random flights as well as their pathwise behavior. Furthermore, in Section \ref{ctrw} we give some remarks on continuous-time random walk (CTRW) approach in this setting. Finally, in Section \ref{ftequation}, we study the particularly interesting one-dimensional case, that is related to the time-fractional telegraph equation widely studied in the mathematical literature. We give a probabilistic interpretation of the solution for the Cauchy problem and  show the relation with the kinetic model equation.

\section{A general approach for random flights in $\mathbb{R}^d$}\label{overview}

In this section we introduce isotropic transport processes and recall their main properties. We consider a $d$-dimensional run-and-tumble walk describing a particle moving at 
constant speed $v$ and changing its direction of motion with rate $\alpha>0,
$ at each collision.  
In particular, after any collision
the particle randomly reorients its direction of motion uniformly on the unit $(d-1)$-dimensional sphere $\mathbb{S}^{d-1}=\{\mathbf{ x}\in\mathbb R^d:||{\bf x}||=1\}$
(see, e.g., \cite{orsingher2007random, martens2012probability,angelani2013averaged}). 
For $d=1$ the above notation means that the new direction is randomly chosen  on 
the discrete set $\mathbb{S}^0=\{-1,1\}$.
The initial direction is  randomly chosen on $\mathbb{S}^{d-1}$.
Let ${p}({\bf x},t;\E)$ be the  probability density function to find the particle 
at position ${\bf x}\in\mathbb R^d$ at time $t$ (for $d=1$ we indicate the position with $x$)  with velocity orientation $\E\in\mathbb{S}^{d-1}$.  
We can write 
the kinetic equation 
for the run-and-tumble motion
as in \cite{stadje1989exact, martens2012probability} (also called forward Kolmogorov equation)
\begin{eqnarray}
\partial_t {p}({\bf x},t;\E)= - v \ \E \cdot \nabla_{\bf x} {p}({\bf x},t;\E)
- \alpha {p}({\bf x},t;\E)
+ \alpha \int_{\mathbb{S}^{d-1}} {p}({\bf x},t;\E')\sigma(d\E') ,
\label{Prte}
\end{eqnarray}
where $\sigma (d {\bf e})=\frac{d\E}{\Omega_d} $ and $\Omega_d= 2 \pi^{d/2} / \Gamma(d/2)$ is the solid angle
in $d$ dimension (i.e. the uniform density law on the $(d-1)$-dimensional unit sphere).
We note that for $d=1,$ the particle moves rightward and leftward and then
we have only two possible directions, that is $\E\in\{-1,1\}$.
The previous equation continues to be valid in $d=1$, bearing in mind that the integral becomes a sum
\begin{equation}
\label{Projd1}
\int_{\mathbb{S}^{d-1}} f(\E) \sigma(d\E)  \to \frac12 \sum_{\E=\pm 1} f(\E) , \ \ \ d=1 ,
\end{equation}
and the  system is described by two hyperbolic equations (see, e.g., \cite{PhysRevE.48.2553} and \cite{weiss2002some})
\begin{eqnarray}\label{eq: ke1}
\partial_t {p}(x,t;\E)= - v \E \cdot \partial_x {p}(x,t;\E)
+ \frac{\alpha}{2} [{p}(x,t;-\E)-{p}( x,t;\E)] .
\label{Prtetp}
\end{eqnarray}
It is worthwhile to observe that the density $p$ is not normalized with respect to orientation $\E;$ that is $\int_{\mathbb{R}^d}\int_{\mathbb{S}^{d-1}}p({\bf x},t;\E)d{\bf x}d\E=\Omega_d,$ where $d\E$ is the surface measure.

Now, we describe the above motions in terms of stochastic processes.
As will become clear later, it is convenient to treat differently the cases  $d\geq2$ and $d=1$. 
For $d\geq2$ 
let $\{N(t): t\geq 0\}$ be a homogeneous Poisson process with rate $\alpha>0$. 
We can describe run-and-tumble motions by means of the  velocity-jump process 
\begin{align}\label{eq:vjp}
  \mathbf{V}(t)={\bf V}_k,\quad T_k\leq t<T_{k+1} ,   
\end{align} where $\{{\bf V}_k: k\geq 0\}$ is a sequence of 
 independent and identically distributed
random variables taking values uniformly on $\mathbb{S}^{d-1}$ (which are independent of $\{N(t): t\geq 0\}$), and $T_k, k\geq 0$ $(T_0=0),$ represent  Poisson jumping times.  The random position reached by the particle at time $t>0$ is denoted by \begin{align}\label{eq:tp}\mathbf{X}(t)=v \int_0^t \mathbf{V}(s)ds
\end{align}
where
$$\int_0^t \mathbf{V}(s)ds=\sum_{k=1}^{N(t)} {\bf V}_{k-1} (T_k-T_{k-1})+{\bf V}_{N(t
)}(t-T_{N(t)}).$$
Therefore, 
we have that for any $A_1\in\mathcal B(\mathbb R^d)$ and  $A_2\in\mathcal B(\mathbb{S}^{d-1}),$ one has that $$\textsc{P}(\mathbf{X}(t)\in A_1, {\bf V}(t)\in A_2)=\iint_{A_1\times A_2}{p}({\bf x},t;\E) d{\bf x} \sigma(d\E).$$
For  $d=1$ it is convenient to define $\{N(t): t\geq 0\}$ as a homogeneous Poisson process 
with rate $\alpha/2$, corresponding to the inversion of the particle velocity instead of 
the simple resetting of its orientation (when we refer to the $d=1$ case in the rest of the 
manuscript we mean exactly this definition of stochastic process).
In this case $$X(t)=v \int_0^t V(s)ds$$ is a telegraph process with $V(t)=V(0)(-1)^{N(t)},$ where $V(0)$ is a random variable assuming values $\pm 1$ with the same probability and independent of $N(t).$

Hereafter, $\{N(t): t\geq 0\}$ stands for a Poisson process with rate $\alpha$ for the random motions in $\mathbb R^d$ with $d\geq2,$ while the rate is fixed as $\alpha/2$ in the one-dimensional case.

\noindent
By introducing the projector operator, defined as an integral 
(sum in $d=1$, see (\ref{Projd1})) over velocity orientations
\begin{equation}
\mathbb{P} f({\bf x},\E)= 
\int_{\mathbb{S}^{d-1}} f({\bf x},\E)\sigma(d\E) , 
\end{equation}
the kinetic equations \eqref{Prte} and \eqref{eq: ke1}, can be put in the form
\begin{eqnarray}\label{eq:adoper}
\partial_t {p}({\bf x},t;\E) = - v \ \E \cdot \nabla_{\bf x} {p}({\bf x},t;\E)
+\alpha (\mathbb{P} - 1)
{p}({\bf x},t;\E) .
\label{Prte2}
\end{eqnarray}
We look for the solution of the equation (\ref{Prte2}) averaged over swimming directions
\begin{equation}
\label{eq:pdfpos}
P({\bf x},t) = \mathbb{P} \ {p}({\bf x},t;\E) ,
\end{equation}
representing the probability density function of the position reached from the particle at time $t.$ Furthermore, we have that 
\begin{align}\label{eq:pdfpos2}
P({\bf x},t) = P_{\text{s}}(t) \delta(||{\bf x}||-vt)+   P_{\text{ac}}({\bf x},t)1_{||{\bf x}||<vt} ,
\end{align}
where the first term represents the singular component of the probability distribution arising when the particle does not change direction up to time $t$, and
the second term is the absolutely continuous component of the probability law of ${\bf X}(t), t>0,$ which lies within $\mathbb S_{vt}^{d-1}$.
The singular term is, in the one-dimensional case
\begin{equation}
\label{Psd1}
P_{\text{s}}(t) =\frac{e^{-\alpha t/2}}{2} ,\ \ \  d=1 ,
\end{equation}
and, in higher dimensions 
\begin{equation}
\label{Psdgeq2}
P_{\text{s}}(t) =\frac{e^{-\alpha t}}{\Omega_d (vt)^{d-1}} ,\ \ \  d\geq2 .
\end{equation}
The $\alpha/2$, instead of $\alpha$, appearing in the exponential for $d=1$ is due to the fact that after a tumble, 
occurring at rate 
 $\alpha$, the particle can proceed along the original direction with probability $1/2$,
 or, in other words, the particle effectively changes (inverts) direction not at the rate
 of tumbling but at its half.

Now, we describe the methodology allowing to explicit the solution $P({\bf x},t)$ in some dimensions. Let $g({\bf x},t)$ be a suitable function; we introduce the Fourier and Laplace transforms, respectively, as \begin{equation*}\hat g({\bf k},t)=\mathcal{F}[g({\bf x},t)]({\bf k},t)=\int_{\mathbb R^d}  \ e^{i {\bf k}\cdot {\bf x}} g({\bf x},t)d{\bf x}, \quad {\bf k}\in \mathbb{R}^d,
\end{equation*} 
and
\begin{equation*}\tilde g({\bf x},s)= \mathcal L[g({\bf x},t)]({\bf x},s)=\int_0^\infty  \ e^{-st}  g({\bf x},t) dt,\quad s\geq 0 .
\end{equation*}
The Fourier-Laplace transform of $ {p}({\bf x},t;\E)$ is denoted by ${\hat {\tilde {p}}} ({\bf k},s;\E)$
and considering the initial condition
\begin{equation}
     {p}({\bf x},0;\E) = 
 {p}_{0}({\bf x};\E) ,
\end{equation}
whose Fourier transform is    ${\hat {p}}_{0} ({\bf k};\E),$
we can write the kinetic equation (\ref{Prte2}) in the Fourier-Laplace domain as
\begin{equation}
\label{RTeqFL}
(s - i v {\bf k} \cdot \E) {\hat {\tilde {p}}} ({\bf k},s;\E) =
\alpha (\mathbb{P}-1) {\hat {\tilde {p}}} ({\bf k},s;\E) + {\hat {p}}_{0} ({\bf k};\E) .
\end{equation}
We specialize to the case in which the  particle starts its motion at the origin with randomly
distributed orientation, 
\begin{equation}
{p}_{0}({\bf x};\E)=\delta ({\bf x}) ,
\end{equation}implying 
${\hat {p}}_{0} ({\bf k};\E) = 1$.
By solving (\ref{RTeqFL}) for ${\hat{\tilde p}}({\bf k},s;\E)$
and applying the projector operator $\mathbb{P}$
we finally arrive at the expression of (\ref{eq:pdfpos}) in the Laplace-Fourier domain
\begin{equation}
{\hat {\tilde P}}({\bf k},s) =
\frac{P_0({\bf k},s)}{1-\alpha P_0({\bf k},s)} = 
\sum_{n=0}^{\infty} \alpha^{n} \ P_0^{n+1}({\bf k},s) ,
\label{PP0}
\end{equation}
where 
\begin{equation}
P_0({\bf k},s) = \mathbb P\left(\frac{1}{s+\alpha - i v {\bf k} \cdot \E}\right) ,
\label{P0}
\end{equation}
and the expansion in \eqref{PP0} is 
justified by 
$|\alpha P_0| < 1$ since (for $|s+\alpha|>\alpha$)
\[
\frac{\alpha}{|s+\alpha - i v {\bf k} \cdot \E|} = \frac{\alpha}{\sqrt{(s+\alpha)^2 + v^2 ({\bf k} \cdot \E)^2}} \leq \frac{\alpha}{|s + \alpha|} < 1. \]

By noting that $P_0$ is a function of $k=||{\bf k}||$ and $s+\alpha,$ we can
write the formal expression of the probability distribution $P(r,t)$ as a function 
of $r=||{\bf x}||$ (therefore the random flights are isotropic) and $t$. Indeed, by passing to the spherical coordinates and using formula (2.12) in \cite{de2012flying}, we get
\begin{eqnarray}
P(r,t) = \frac{1}{r^{\frac{d}{2}-1}} \sum_{n=0}^{\infty} \alpha^{n} \int_0^\infty \frac{dk}{(2\pi)^{\frac{d}{2}}} 
 k^{\frac{d}{2}} \ J_{\frac{d}{2}-1}(kr) \
 {\mathcal L}^{-1}[ P_0^{n+1}(k,s)](k,t) ,
\label{Prt}
\end{eqnarray}
with $J_\nu(x)=\sum_{k=0}^\infty (-1)^k\frac{(x/2)^{2k+\nu}}{\Gamma(k+\nu+1)},x,\nu\in \mathbb R,$  the Bessel function of the first kind and ${\mathcal L}^{-1}$ the inverse Laplace transform. 
For completeness we also report the expressions of the PDF in Fourier and Laplace domains
\begin{equation*}
{\hat P}(k,t) = \sum_{n=0}^{\infty} \alpha^{n} 
 {\mathcal L}^{-1}[ P_0^{n+1}(k,s)](k,t) ,
\label{Prt1}
\end{equation*}
and
\begin{equation*}
{\tilde P}(r,s) = \frac{1}{r^{\frac{d}{2}-1}} \sum_{n=0}^{\infty} \alpha^{n} \int_0^\infty \frac{dk}{(2\pi)^{\frac{d}{2}}} 
 k^{\frac{d}{2}} \ J_{\frac{d}{2}-1}(kr) \ P_0^{n+1}(k,s) .
\label{Prt2}
\end{equation*}
Explicit expressions of $P(r,t)$ can be obtained when 
one is able to explicitly invert the Laplace transform of $P_0^{n+1}$, 
calculate the integral on $k$ and sum the series.
This is, for example , the case of $d=1$ and $2$.
In the one-dimensional case one has
\begin{eqnarray}
P(x,t) & = & e^{-\alpha t/2} \left\{
\frac{\delta(x-vt)+\delta(x+vt)}{2}  \right. \nonumber \\
 & +& \left.
\frac{\alpha}{4v}
\left[
I_0(\alpha \Delta /2v) +\frac{vt}{\Delta} I_1(\alpha \Delta/2v)
\right] \theta(vt-|x|)
\right\} ,
\label{PDF1d}
\end{eqnarray}
where $\Delta=\sqrt{v^2t^2-x^2}$
and $I_\nu(x)=\sum_{k=0}^\infty \frac{(x/2)^{2k+\nu}}{\Gamma(\nu+k+1)}, x,\nu\in \mathbb{R},$ is the modified Bessel function and $\theta$ represents the Heaviside function (see, e.g., \cite{weiss2002some} and \cite{martens2012probability}). Furthermore, it is well-known that the telegraph process is linked to the telegraph hyperbolic equation (also called damped wave equation), since $P(x,t)$ is the fundamental solution of the Cauchy problem (see, e.g., \cite{weiss2002some})
\begin{align}\label{eq:cpt}
    &\partial_t^2u(x,t)+ \alpha
    \partial_tu(x,t)=v^2\partial_{xx}^2 u(x,t),\\
    &u(x,0)=\delta(x),\quad \partial_tu(x,0)=0 .\notag
\end{align}
Furthermore if we replace in \eqref{eq:cpt} the initial condition with $u(x,0)=\phi(x),$ where $\phi\in C^2,$ we obtain the interesting stochastic solution derived in \cite{kac1974stochastic}
\begin{align}\label{eq:kss}
u(x,t)=\frac12\left(\mathbb E\left[\phi\left(x-v\int_0^t (-1)^{N(s)}ds\right)\right]+\mathbb E\left[\phi\left(x+v\int_0^t (-1)^{N(s)}ds\right)\right]\right) .
\end{align}

In two-dimensions we have (see, e.g., \cite{monin1956statistical, stadje1987exact, kolesnik2005planar, martens2012probability})
\begin{equation}
	P(r,t) = e^{-\alpha t} \left[
	\frac{\delta(r-vt)}{2\pi r} +
	\frac{\alpha}{2 \pi  v \Delta} \exp{\left(\frac{\alpha \Delta}{v}\right)}
	\theta(vt-r) \right] ,
  \label{Prt2d}
\end{equation}
with $\Delta=\sqrt{v^2t^2-r^2}$.
It is worth noting that the case $d=3$ has not explicit solution, while,
interestingly, $d=4$ does 
(see \cite{Paas1997,orsingher2007random,Detch2014}). 

It is useful to describe some features of the random motions by means of the mean square displacement (MSD), that can be easily calculated as 
\begin{equation}
{\bf r}^2(t) = \int_{\mathbb R^d} r^2 P(r,t)  d{\bf x} = - \left. \nabla^2_{\bf k} \hat{P}({\bf k},t) \right|_{{\bf k}\!=\!0} .
\label{r2P}
\end{equation}
In the Laplace domain, from (\ref{PP0}) we obtain
\begin{align*}
\nabla_{\bf k} {\hat {\tilde P}} &=  \frac{\nabla_{\bf k} P_0}{(1-\alpha P_0)^2} , \\
\nabla^2_{\bf k} {\hat {\tilde P}} &=   \frac{\nabla^2_{\bf k} P_0}{(1-\alpha P_0)^2} 
+ \frac{2\alpha}{(1-\alpha P_0)^3}  (\nabla_{\bf k} P_0)\cdot(\nabla_{\bf k} P_0) .
\end{align*}
From (\ref{P0}) and its derivatives we have that, at ${\bf k}=0$,
\begin{align*}
P_0|_{{\bf k}\!=\!0} &= \frac{1}{s+\alpha} , \\
\left. \nabla_{\bf k} P_0\right|_{{\bf k}\!=\!0} &=0  , \\
\nabla^2_{\bf k} P_0|_{{\bf k}\!=\!0} &= -\frac{2v^2}{(s+\alpha)^{3}} ,
\end{align*}
which allows us to obtain the expression of the MSD in the Laplace domain
\begin{equation}
\widetilde{{\bf r}^2}(s) = \frac{2v^2}{s^2(s+\alpha)} .
\label{r2L}
\end{equation}
Rewriting the above expression in the form 
$$
\widetilde{{\bf r}^2}(s) = \frac{2v^2}{\alpha^2} \left[ \frac{\alpha}{s^2} -\frac{1}{s} +\frac{1}{s+\alpha} \right],
$$
and considering inverse Laplace transforms \cite{gradshteyn2014table},
we finally obtain the MSD in the time domain
\begin{equation}
{\bf r}^2(t) = \frac{2v^2}{\alpha^2} (\alpha t -1 + e^{-\alpha t}) .
\label{r2t}
\end{equation}
It is worth noting that, although the form of PDF (and also the existence of an explicit expression for it) 
depends on the dimension of the scattering environment, its second moment is independent of $d$.

For the random flights  it is also useful to deal with the backward Kolmogorov equation 
\begin{align}\label{eq:cauchy}
\partial_t u({\bf x}, t; {\bf e})=\mathfrak L     u({\bf x}, t; {\bf e}), \quad  u({\bf x},0 ; {\bf e})=f({\bf x}, {\bf e}) ,
\end{align}
where 
\begin{align}\label{eq:igrf}
    	{\mathfrak L} :=  v \ \E \cdot \nabla_{\bf x} 
+\alpha (\mathbb{P} - 1)
\end{align}
is the infinitesimal generator of the strong Markov process $\{({\bf x}+{\bf X}(t),{\bf V}(t)): t\geq 0\},$ and $f\in$ Dom$(\mathfrak L)=\{f\in L^2(\mathbb R^d\times \mathbb S^{d-1}):v \ \E \cdot \nabla_{\bf x}f\in  L^2(\mathbb R^d\times \mathbb S^{d-1}) \}.$ It is worth to mention that the operator appearing on the right side of \eqref{eq:adoper}, that is $-v \ \E \cdot \nabla_{\bf x} 
+\alpha (\mathbb{P} - 1),$ represents the adjoint of ${\mathfrak L}.$ Therefore, the unique solution of Cauchy problem \eqref{eq:cauchy} admits the following stochastic interpretation
\begin{align}
   u({\bf x},t; {\bf e})=\mathbb E_{\bf e} f({\bf x}+{\bf X} (t), {\bf V} (t)) ,
\end{align}
 given the starting position and direction $({\bf x},\E)\in \mathbb R^d\times \mathbb S^{d-1}$ of the particle ($\mathbb E_{\bf e}$ stands for the mean conditionally on ${\bf V}(0)={\bf e}$). Furthermore, for $d\geq 2$ 
 \begin{align}\label{eq:integeqrf}
  u({\bf x}, t;{\bf e})&=f({\bf x}+v {\bf e} t, {\bf e}) e^{-\alpha t}+\alpha\int_0^{t}e^{-\alpha s} \int_{\mathbb S^{d-1}} u(t-s, {\bf x}+v{\bf e}s;\E')\sigma(d\E')ds
 \end{align}
(see, e.g., Lemma 2.1 in \cite{watanabe1970convergence}), while if $d=1$
 \begin{align*}
  u(x, t;{\bf e})&=f(x+v {\bf e} t, {\bf e}) e^{-\frac{\alpha}{2} t}+\frac{\alpha}{2}\int_0^{t}e^{-\frac{\alpha}{2} s} u(t-s,  x+v{\bf e}s;-\E)ds .
 \end{align*}


\section{Time-fractional kinetic equations} \label{fkequation}

The main idea of this paper is to introduce a fractional version of the classical kinetic equation \eqref{Prte} and then to analyze the related random model. 
While there is a wide literature about the fractional telegraph equation, we underline that, as far as we know, there is not a general theory regarding the modified kinetic equation obtained by replacing the ordinary time derivative with the Caputo derivative of order $\nu\in (0,1)$. 
The main object of the paper is to provide a complete analysis to this general exploratory mathematical problem in order to understand, \textit{a posteriori}, its meaning and the potential utility in physical and probabilistic models. 
The strong motivation for this study is given by the great interest in the physical models for the time-fractional Cattaneo equation that is just a particular, but relevant, special case of the general fractional kinetic equation that we are going to discuss in detail. In our view all this suggests that a general theory is fundamental to better clarify the non-trivial impact of the time-fractional generalization. Heuristically the time-fractional generalization of kinetic models leads to anomalous diffusion, due to the so-called \textit{memory effects}. Here we provide the rigorous probabilistic interpretation of the related processes starting from the general $d$-dimensional fractional kinetic equation.

Let us start by introducing the time-fractional kinetic equation in space dimension $d$ given by
\begin{equation}\label{eq:fked}
\partial_t^{\nu} {p}_\nu({\bf x},t;\E) = - v \ \E \cdot \nabla_{\bf x} {p}_\nu({\bf x},t;\E)
+\alpha (\mathbb{P} - 1)
{p}_\nu({\bf x},t;\E) ,
\end{equation}
where $({\bf x}, \E)\in \mathbb R^d\times \mathbb S^{d-1},$ and $p_{0,\nu}({\bf x},\E)=p_\nu({\bf x},0;\E).$  The standard time derivative appearing in \eqref{Prte} has been replaced with the Caputo time-fractional derivative \eqref{de:fd} of order $\nu \in (0,1)$.
Clearly for $\nu=1$ the equation \eqref{eq:fked} reduces to  \eqref{Prte}.
Strictly speaking, for dimensional reasons, we should introduce a 
factor $\tau_0^{\nu-1}$ on the left side of the previous equation,
with $\tau_0$ an arbitrary time-scale parameter. In the following, without loss of generality, 
we express times in unit of $\tau_0$, i.e., we set $\tau_0=1$.
In the Fourier-Laplace domain the equation \eqref{eq:fked} becomes
\begin{equation*}
(s^\nu - i v {\bf k} \cdot \E) {\hat {\tilde {p}}}_\nu ({\bf k},s;\E) =
\alpha (\mathbb{P}-1) {\hat {\tilde {p}}}_\nu ({\bf k},s;\E) + 
s^{\nu-1}{\hat {p}}_{0,\nu} ({\bf k};\E) ,
\end{equation*}
having used the property \eqref{ltfd} of the Laplace transform of the Caputo derivative.
Proceeding as before, we can obtain the averaged probability density function
$P_\nu({\bf x},t) = \mathbb{P} \ {p}_\nu({\bf x},t;\E)$, valid for initial condition 
${p}_{0,\nu}({\bf x};\E)=\delta ({\bf x})$,
\begin{equation}
{\hat {\tilde P}}_\nu({\bf k},s) = s^{\nu-1}
\frac{P_0({\bf k},s)}{1-\alpha P_0({\bf k},s)}  ,
\label{Pks_fr}
\end{equation}
where 
\begin{equation}
\label{P0nu}
P_0({\bf k},s) = \mathbb{P}\left( \frac{1}{s^\nu+\alpha - i v {\bf k} \cdot \E}\right) .
\end{equation}
Therefore, the above expressions allow us to express  $P_\nu$
in terms of the classical probability density function $P$ investigated in the previous section:
\begin{equation*}
 \hat {\tilde{P}}_\nu({\bf k},s) = s^{\nu-1}
{\hat{\tilde{P}}}({\bf k},s^\nu) ,
\end{equation*}
or, in the variable ${\bf x}$,
\begin{equation}\label{eq:ltfdf}
\tilde {P}_\nu({\bf x},s) = s^{\nu-1}
{\tilde P} ({\bf x},s^\nu) .
\end{equation}
We show that $P_\nu$ represents itself a probability density function. Indeed, from \eqref{eq:ltfdf} by integrating with respect to variable ${\bf x}$, we derive
\begin{align*}
    \int_0^\infty e^{-st}\left(\int_{\mathbb R^d} P_\nu({\bf x},t) d {\bf x}  \right)dt&=s^{\nu-1} \int_0^\infty e^{-s^\nu t}\left(\int_{\mathbb R^d} P({\bf x},t) d {\bf x}  \right)dt\\
    &=s^{\nu-1} \int_0^\infty e^{-s^\nu t}dt\\
    &=\frac1s .
\end{align*}
Therefore the above equality holds if and only if $\int_{\mathbb R^d} P_\nu({\bf x},t) d {\bf x} =1.$ 
The non-negativity of $P_\nu({\bf x}, t)$ 
follows from the fact that $\tilde P_\nu({\bf x}, s)$ can be expressed as a product of two completely monotone (CM) functions, as in \eqref{eq:ltfdf}. Recall that an infinitely differentiable  function $f(s)$ is said to be completely monotone if 
$
(-1)^n f^{(n)}(s) \geq 0
$
for all $s > 0$ and non-negative integer $n$, whereas it is said to be a Bernstein function
if 
$
(-1)^{n-1} f^{(n)}(s) \geq 0
$ for all $s > 0$ and $n \in \mathbb N$.
It is immediate to check that $u(s) = s^{\nu -1}$
is completely monotone. By Bernstein's theorem \cite{schilling-bern}, Theorem 1.4, also 
$v(s) = {\tilde P} ({\bf x},s)$ is. By Theorem 3.7 in \cite{schilling-bern}, $s\mapsto {\tilde P} ({\bf x},s^\nu)$ is CM since it is the composition of the CM function $v$ and the Bernstein function $s \mapsto s^\nu$.
Since the product of CM functions is easily seen to be CM, see e.g. Corollary 1.6 in \cite{schilling-bern}, by Bernstein's theorem  
$\tilde P_\nu$ is the Laplace transform of a measure. The conclusion follows by the uniqueness of the Laplace transform. Therefore $P_\nu$ represents a density function and then the equation \eqref{eq:fked} describes a random motion.

All the results obtained in the previous section can then be used to 
obtain the Laplace transformed PDFs in the case of fractional derivative processes.

The MSD ${\bf r}_\nu^2=\int_{\mathbb R^d} r^2P_\nu(r,t) d{\bf x} $ associated to $P_\nu$ can be calculated using (\ref{r2P})
obtaining, in the Laplace domain,
\begin{equation}
\widetilde{{\bf r}_\nu^2}(s) = \frac{2v^2}{s^{\nu+1}(s^\nu+\alpha)}  ,
\label{r2fracL}
\end{equation}
in agreement with the one-dimensional expression reported in \cite{angelani2020fractional}.
The inverse Laplace transform of equation \eqref{r2fracL} provides 
the explicit form of the MSD regardless of dimension $d$, i.e. (see (\ref{LML}) in Appendix \ref{App2})
\begin{equation}\label{r2}
	{\bf r}_\nu^2(t) = 2v^2 t^{2\nu}E_{\nu,2\nu+1}(-\alpha t^\nu) ,
\end{equation}
in agreement with equation (37) in \cite{sevilla2023anomalous}. 
We recall that the function $E_{\nu,2\nu+1}(-\alpha t^\nu)$ appearing in \eqref{r2} is the well-known two-parameter Mittag-Leffler function, whose general form can be expressed in series form as
\begin{equation}
	E_{\alpha,\beta}(z) = \sum_{n=0}^\infty \frac{z^n}{\Gamma(n\alpha+\beta)},\quad z\in\mathbb C, \alpha,\beta>0 ,
\end{equation}
and $E_{\alpha,1}(z)=E_\alpha(z).$
It can be proved by simple calculations that for $\nu= 1$ we recover the MSD \eqref{r2t} of the classical telegraph process (we refer to \cite{gorenflo2020mittag} for the properties of the two parameter Mittag-Leffler function).

The long and short time behavior of the MSD can be obtained from the asymptotic 
form of its Laplace transform (\ref{r2fracL}), by using Tauberian theorems 
(see, e.g., Theorem 2-3, Chapter XIII, \cite{feller1991introduction}). 
According to this theory the asymptotic behaviour of the function ${\bf r}_\nu^2(t)$ for $t \rightarrow + \infty$ and $t \rightarrow 0$ can be formally obtained from the asymptotic behaviour of its Laplace transform respectively for $s\rightarrow 0^+$ and for $s\rightarrow +\infty$.
Therefore, from \eqref{r2fracL} we can easily prove that at short time the MSD behaves as
\begin{equation}\label{eq:fmsd1}
    {\bf r}_\nu^2(t) \sim \frac{2v^2}{\Gamma(2\nu+1)} \ t^{2\nu} ,
    \hspace{1cm} t \to 0 .
\end{equation}
while, in the long time limit, we have
\begin{equation}\label{eq:fmsd2}
    {\bf r}_\nu^2(t) \sim \frac{2v^2}{\alpha \Gamma(\nu+1)}\ t^\nu ,
    \hspace{1cm} t \to \infty .
\end{equation}
The results \eqref{eq:fmsd1} and \eqref{eq:fmsd2} reveal that, as expected, the scattering random motion governed by the fractional kinetic equation \eqref{eq:fked} has an anomalous behavior, because the asymptotic MSD is not linear in time, but of order 
$t^\nu.$   

\section{Anomalous transport processes with random time}\label{atprocesses}
In order to give a stochastic interpretation of the solution to  the equation \eqref{eq:fked}, let us consider the time-fractional version of the  Cauchy problem \eqref{eq:cauchy} given by
	\begin{equation}\label{fra}
	\partial_t^{\nu} u_\nu({\bf x},t;\E) ={\mathfrak L}
		u_\nu({\bf x},t;\E),\quad  u_\nu({\bf x},0;\E)=f({\bf x},\E) ,
	\end{equation}
	where ${\mathfrak L} $ 
 is the infinitesimal generator \eqref{eq:igrf} of the couple $({\bf x}+{\bf X}(t),{\bf V}(t)), t\geq 0,$ and $f\in$Dom$(\mathfrak L).$ We resort to the general theory developed in \cite{baeumer2001stochastic, meerschaert2004limit} on the time-fractional abstract Cauchy problem and its stochastic solution defined by means of time-changed Markov processes (see Appendix \ref{App1}).
 
 First of all, we recall that a subordinator is a non-negative and non-decreasing Lévy process starting from zero (see, e.g., \cite{applebaum2009levy}). A stable subordinator 
 $\{L_\nu(u)$: $u\geq 0\}$
 is a strictly increasing L\'evy subordinator with Laplace exponent given by
 \begin{equation}\label{eq:levy-sym}
\mathbb{E}\left[e^{-sL_\nu(u)}\right]= e^{-u s^\nu }, \quad \nu \in(0,1).
 \end{equation}
 We denote with $g_\nu$ the probability density function of
 the stable L\'evy subordinator $L_\nu(u)$.
The inverse stable subordinator $\{Y_\nu(t):t\geq 0\}$ (see, e.g., \cite{meerschaert2019inverse}) 
 \begin{equation}
    Y_\nu(t)= \inf\{u>0: L_\nu(u)>t\}, \quad t >0 ,
 \end{equation}
 with $Y_\nu(0)=0$ a.s., is such that
 \begin{equation}\label{eq:tlis}
    \mathbb{E}\left[e^{-sY_\nu(t)}\right]=E_\nu(-st^\nu) .
 \end{equation}
 The inverse process is a non-Markovian with non-stationary, non-independent increments and non-decreasing continuous a.s. sample paths.  
 The probability density function of $Y_\nu(t), t>0,$ is given by $\mu_\nu$
(see Appendices \ref{App1} and \ref{App2}).
 
Let $\{N(Y_\nu(t)):t\geq 0\}$ be the fractional Poisson process obtained time-changing the classical Poisson process $N(t)$ with $Y_\nu(t).$ This process coincides with a renewal process with i.i.d. waiting times between two consecutive jumps given by $\{J_n: n\in\mathbb N\}$  with $P(J_n>t)=E_\nu(-\alpha t^\nu),$ for $d\geq 2$ and  $P(J_n>t)=E_\nu(-\frac\alpha2 t^\nu),$ for $d=1$ (see \cite{10.1214/EJP.v16-920} 
 for more  details on this process). We observe that $ L_\nu(T_k-)=\sup\{t>0: Y_\nu(t)<T_k\}$ coincides with the $k$-th jumping time $J_1+...+J_k$ of $\{N(Y_\nu(t)):t\geq 0\}$ (see Lemma 2.1 and Theorem 2.2 in \cite{10.1214/EJP.v16-920}). 
 
 By exploiting the general theory recalled in Appendix which applies to our case, we can  claim that the stochastic process governed by Eq. (\ref{fra}) corresponds to a time-change with the inverse of the stable subordinator $Y_\nu(t)$ (see \eqref{eq:ssafcp}). Therefore,
the unique solution of the problem \eqref{fra} is given by
\begin{equation}\label{eq:ssfra}
u_\nu({\bf x},t; {\bf e})=\mathbb E_{\bf e} f({\bf x}+{\bf X} (Y_\nu(t)), {\bf V} (Y_\nu(t))) .
\end{equation}

	 This means that the fractional equation (\ref{fra}) is the governing equation for the  couple $$\{({\bf x}+{\bf X} (Y_\nu(t)), {\bf V} (Y_\nu(t))):t\geq 0\},$$
where the original processes \eqref{eq:vjp} and \eqref{eq:tp}   are deformed by a random clock. The time-changed jump-velocity process becomes
\begin{align}
    {\bf V} (Y_\nu(t))&= {\bf V}_k,\quad T_k\leq Y_\nu(t) < T_{k+1} ,
\end{align}
or equivalently
\begin{align}
    {\bf V} (Y_\nu(t))=
    {\bf V}_k,\quad L_\nu(T_k-)\leq t < L_\nu(T_{k+1}-) .
\end{align}
The time-changed random flight is given by
\begin{align}\label{eq:rptc}
   \mathbf{X}(Y_\nu(t))&=v\int_0^{Y_\nu(t)} \mathbf{V}(s)ds\\
   &=\begin{cases}
       v {\bf V}_0Y_\nu(t),& 0\leq Y_\nu(t)<T_1,\\
         \mathbf{X}(T_1)+ v{\bf V}_1( Y_\nu(t)-T_1),& T_1\leq Y_\nu(t)<T_2\\ 
        ...\notag\\
        \mathbf{X}(T_{N(Y_\nu(t))})+ v{\bf V}_{N(Y_\nu(t))}( Y_\nu(t)-T_{N(Y_\nu(t))}),&  Y_\nu(t)\geq T_{N(Y_\nu(t))}
   \end{cases}\notag\\
   &=\begin{cases}
       v {\bf V}_0Y_\nu(t),& 0\leq t<L(T_1-),\\
         \mathbf{X}(T_1)+ v{\bf V}_1( Y_\nu(t)-T_1),& L(T_1-)\leq t<L(T_2-)\\ 
        ...\notag\\
        \mathbf{X}(T_{N(Y_\nu(t))})+ v{\bf V}_{N(Y_\nu(t))}( Y_\nu(t)-T_{N(Y_\nu(t))}),&  t\geq L(T_{N(Y_\nu(t))}-)
   \end{cases}\notag\\
   &=v\sum_{k=1}^{N(Y_\nu(t))} \mathbf{V}_{k-1} (T_k-T_{k-1})+v\mathbf{V}_{N(Y_\nu(t)
)}(Y_\nu(t)-T_{N(Y_\nu(t))})\notag ,
\end{align}
where clearly we assume that $\sum_{k=1}^0=0$.
In particular, for $d=1$, we obtain the time-changed telegraph process
\begin{align}\label{eq:tctp}
   X(Y_\nu(t))=vV(0)\sum_{k=1}^{N(Y_\nu(t))} (-1)^{k-1} (T_k-T_{k-1})+vV(0)(-1)^{N(Y_\nu(t)
)}(Y_\nu(t)-T_{N(Y_\nu(t))}).
\end{align}
Furthermore, in this last case $(d=1)$, by recalling that (see \cite{pinsky1991lectures})
$$\textsc{P}(V(t)=\pm v|V(0)=\pm v)=\textsc{P}(N(t)\,\, \text{even})=\frac12(1+e^{-{\alpha} t}),$$
$$\textsc{P}(V(t)=\pm v|V(0)=\mp v)=\textsc{P}(N(t)\,\, \text{odd})=\frac12(1-e^{-{\alpha} t}),$$
from \eqref{eq:tlis} follows that the time-changed velocity jumping process ${\bf V}(Y_\nu(t))=V(0)(-1)^{N(Y_\nu(t))}$ representing a semi-Markov chain with
$$\textsc{P}(V(Y_\nu(t))=\pm v|V(0)=\pm v)=\frac12(1+E_\nu(-{\alpha} t^\nu)),$$
$$\textsc{P}(V(Y_\nu(t))=\pm v|V(0)=\mp v)=\frac12(1-E_\nu(-{\alpha} t^\nu)) .$$

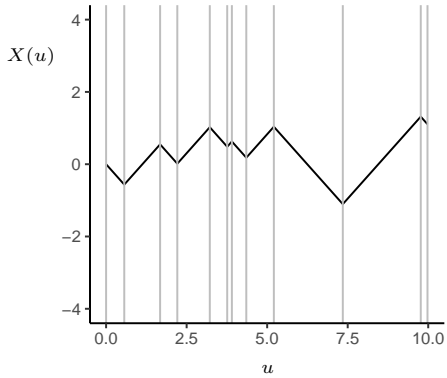
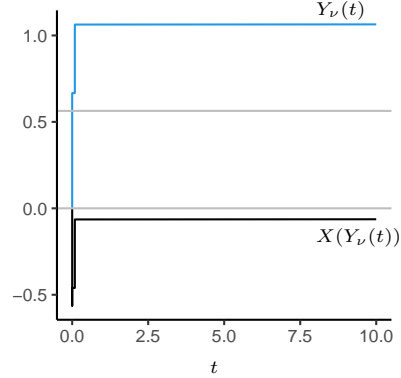
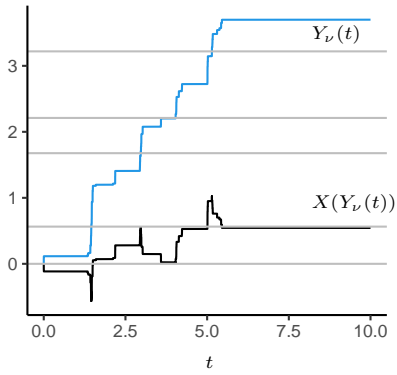
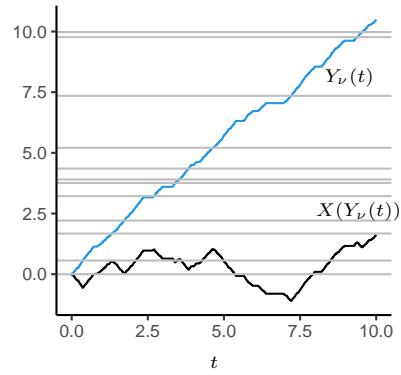
\begin{figure}
        \centering
        \begin{subfigure}[b]{0.475\textwidth}
            \centering
        \captionsetup{skip=10pt}
            \begin{overpic}[width=.8\textwidth]{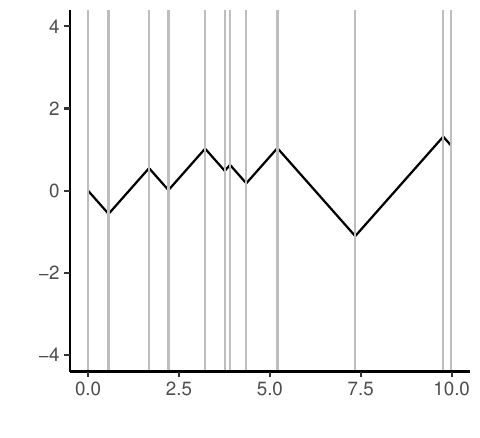}
            \put(-5,76){\scriptsize $X(u)$}
            \put(55,2){\scriptsize $u$}
            \end{overpic}
            \caption{\small ${X}$ ($\alpha=1, v=1$).} 
        \end{subfigure}
        \begin{subfigure}[b]{0.475\textwidth}  
            \captionsetup{skip=10pt}
            \centering 
            \begin{overpic}[width=0.8\linewidth]{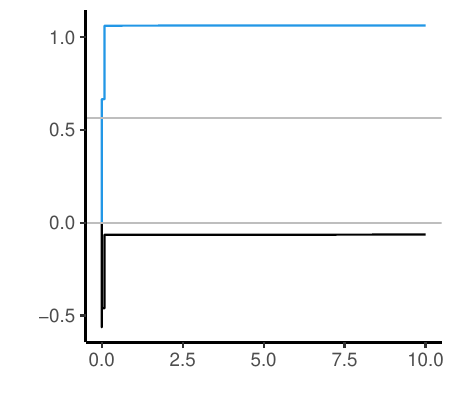}
            \put(80,33){\scriptsize $X(Y_\nu(t))$}
            \put(80, 87){\scriptsize $Y_\nu(t)$}
            \put(55,2){\scriptsize $t$}
            \end{overpic}
            \caption[]%
            {\small $X(Y_\nu)$ , $Y_\nu$ ($\nu$=0.1).}    
        \end{subfigure}
        \vskip\baselineskip
        \begin{subfigure}[b]{0.475\textwidth}   
            \captionsetup{skip=10pt}
            \centering 
            \begin{overpic}[width=0.8\linewidth]{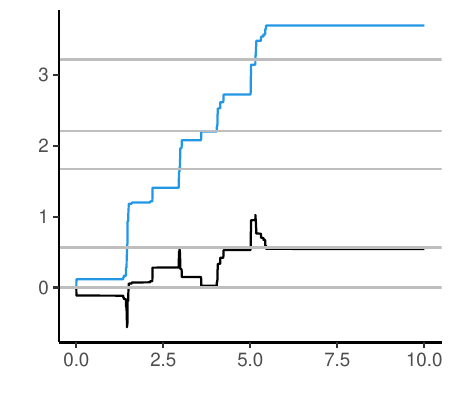}
            \put(80,40){\scriptsize $X(Y_\nu(t))$}
            \put(80, 80){\scriptsize $Y_\nu(t)$}
            \put(55,2){\scriptsize $t$}
            \end{overpic}
            \caption[]%
            {{\small\small ${X}(Y_\nu)$ , $Y_\nu$ ($\nu$=0.5).}}    
        \end{subfigure}
        \begin{subfigure}[b]{0.475\textwidth}   
            \captionsetup{skip=10pt}
            \centering 
            \begin{overpic}[width=0.8\linewidth]{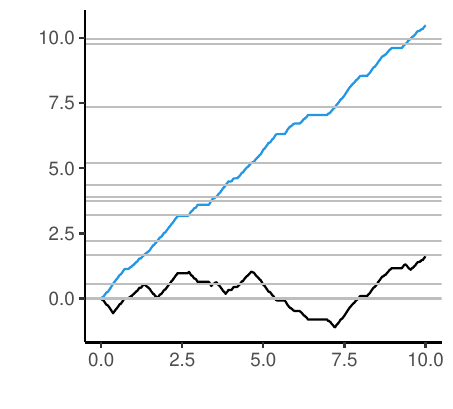}
            \put(80,38){\scriptsize $X(Y_\nu(t))$}
            \put(82, 70){\scriptsize $Y_\nu(t)$}
            \put(55,2){\scriptsize $t$}
            \end{overpic}
            \caption[]%
            {{\small \small ${X}(Y_\nu)$ , $Y_\nu$ ($\nu$=0.9).}}    
        \end{subfigure}
        \caption[]
        {Simulated paths of  ${X}$, ${X}(Y_\nu)$ (black), $Y_\nu$ (blue). The grey horizontal/vertical lines represent the velocity change times $T_k$. $Y_\nu$ is the inverse to a subordinator with density $g_\nu$.} 
        \label{fig:tchange-path}
\end{figure}

Simulated sample paths of process
\eqref{eq:tctp} are shown in \autoref{fig:tchange-path}. 
Panel (A) shows a trajectory of the classical telegraph process $X$ with $v=\alpha=1$. Panels (B)-(C)-(D) show
the sample paths of ${X}(Y_\nu)$ for $\nu = 0.1, 0.5, 0.9,$ respectively. The same initial sample path of $X$ has been used in each plot, while the subordinators have been  independently simulated.
We note that, as a consequence of the construction in \eqref{eq:tctp}, the sample path of ${X}(Y_\nu)$ (black line) 
corresponds to the  juxtaposition of shifted and reflected pieces of the path of  $Y_\nu$. Specifically, whenever the path $Y_\nu$ (blue line) crosses a velocity change random time $T_k$ (horizontal grey lines), the path 
of $X(Y_\nu)$ undergoes a change in direction, as described in \eqref{eq:rptc}. The sample paths of 
$Y_\nu$ have been simulated starting from realizations of a Lévy subordinator $L_\nu$ on a discretized time grid with $\Delta t = 10^{-3}$. 
The pictures display different behaviours of the sample paths of $X(Y_\nu)$ and $Y_\nu$ as $\nu$ varies. Indeed, for small values of $ \nu$ (\autoref{fig:tchange-path} (B)), the subordinator $L_\nu$ admits big jumps (corresponding to the long trapping effect in the sample path) and  many infinitely small jumps. As $\nu$ approaches 1 (\autoref{fig:tchange-path} (D)),
 the subordinator tends to coincide with the deterministic time $t$ and the time-changed sample path shows closer resemblance to the original path of $X$.  In this case the picture shows that ${X}(Y_\nu)$ remains constant for small time intervals. In \autoref{fig:tchange-path} (C) is represented  an intermediate behavior of the time-changed process compared to the Panels (B) and (D).

The probability law of $ \mathbf{X}(Y_\nu(t)), t>0,$ is obtained by averaging the density function of the original process with respect to the probability law of the  random time-change $Y_\nu(t);$ i.e.

\begin{equation}\label{eq:inte}
\int_0^\infty P({\bf x},u)\mu_\nu(u,t)du .
\end{equation}
Now, we prove that the probability density function $P_\nu,$ obtained from the solution of \eqref{eq:fked}, coincides with \eqref{eq:inte}; i.e.
\begin{equation}
\label{PnuP}
P_\nu({\bf x},t)=\int_0^\infty P({\bf x},u)\mu_\nu(u,t)du .
\end{equation}
The above equation is exactly the one obtained in the previous section
in the Laplace domain, eq. (\ref{eq:ltfdf}). 
Indeed, by noting that the Laplace transform of the pdf $g_\nu(u,t)$ of the stable 
subordinator $L_\nu(t)$ with respect to the variable $u$ is 
${\tilde g}_\nu(s,t) =e^{-t s^\nu}$
(see \cite{martens2012probability}),
we have that the Laplace transform of the pdf $\mu_\nu(u,t)$  with respect to $t$ is
${\tilde \mu}_\nu (u,s) = s^{\nu-1} e^{-u s^\nu}$ (see Appendix \ref{App2}).
Therefore, the Laplace transform of (\ref{PnuP}) reads
\begin{equation}
    {\tilde P}_\nu({\bf x},s) = \int_0^\infty P({\bf x},u) {\tilde \mu}_\nu(u,s)du  =
    s^{\nu-1} {\tilde P}({\bf x},s^\nu) ,
\end{equation}
which is exactly the eq. (\ref{eq:ltfdf}).

The equation \eqref{PnuP} allows us to formally write the solution of the time-fractional kinetic
equation as a  superposition of solutions of the classical (non-fractional) equation evaluated 
at all times $u$ and weighted with the (time-dependent)  pdf $\mu_\nu(u,t)$.
We note that for $\nu \to 1$ we recover the classical case, as $\lim_{\nu\to1} \mu_\nu(u,t) = \delta(t-u)$.
For generic $\nu<1$ the pdf $\mu_\nu(u,t)$ has support in $(0,+\infty)$ in the $u$ variable
for any $t>0$, thus allowing the particle to be at any arbitrary distance at any given time $t$ with positive probability; that is given $M>0$
\begin{align*}
\textsc{P}(||{\bf X}(Y_\nu(t))||>M)\leq \textsc{P}(Y_\nu(t)>M/v)=\textsc{P}(L_\nu(M/v)<t)=\int_0^t g(w,M/v)d w ,
\end{align*}
for each $t>0.$
Indeed,  the singular component appearing in $P({\bf x},t)$ is spread over $\mathbb R^{d};$ that is from \eqref{eq:pdfpos2} and \eqref{PnuP} we get for any ${\bf x}\in\mathbb R^{d}$
\begin{align*}
 P_\nu({\bf x},t)= 
 P_{\text{s}}( ||{\bf x}||/v)
 \mu_\nu(||{\bf x}||/v,t)+  \int_0^\infty  P_{\text{ac}}({\bf x},u)1_{||{\bf x}||<vu} \mu_\nu(u,t)d u ,
\end{align*}
where $P_{\text s}(t)$ is given by (\ref{Psd1}) and (\ref{Psdgeq2}).
In other words, the  underlying process is then no longer associated to a finite velocity
random motion.
To further clarify this point we give an alternative representation of the equation (\ref{PnuP}).
Let us first explicitly indicate the  dependence of the PDF on the parameters $\alpha$ and $v$,
as $P({\bf x},t;\alpha,v)$. 
The classical solution $P$ has the following scaling property
\begin{equation}
P({\bf x},u t;\alpha,v) = P({\bf x},t;u \alpha, u v) ,
\end{equation}
as easily obtained from the scaling of (\ref{PP0}),
${\tilde{\hat P}}({\bf k},s/u;\alpha,v) = u {\tilde{\hat P}}({\bf k},s;u \alpha,u v)$,
and the property of the Laplace transform 
${\mathcal L }[f(ut)](s) =u^{-1} {\mathcal L }[f(t)](s/u) $.
Using such a property and making a change of integration variable in (\ref{PnuP}), 
$u \to t^\nu u$, we can finally obtain the following alternative form of $P_\nu$
in term of $P$
\begin{equation}
\label{PnuP2}
P_\nu({\bf x},t;\alpha,v)=\int_0^\infty P({\bf x},t^\nu;u \alpha,u v) \ \mu_\nu(u,1)du .
\end{equation} 
We then conclude that the PDF of the time-fractional process can be viewed as a superposition of 
classical PDFs at rescaled time $t^\nu$ averaged over different tumbling rate $\alpha$ 
and speed $v$ (with constant persistent length $\ell=v/\alpha$) weighted 
with the (time-independent) pdf $\mu_\nu$. 
This clarifies why the finite velocity property is lost in the fractional case.

It is worth mentioning that from the representation \eqref{eq:rptc}, the sample paths of the process $\mathbf{X}(Y_\nu(t)), t\geq 0,$ show an anomalous behavior, while in the classical case the trajectories of the particle are represented by straight lines. The random time change leads to a non-linear dependence with respect to the time of the sample paths of the process; besides the particle is trapped in the same position when $Y_\nu(t)$ is constant (see \autoref{fig:tchange-path}).

Furthermore, for $d\ge 2,$ from \eqref{eq:integeqrf} we get that $u_\nu({\bf x}, t;{\bf e})$  satisfies the following integral equation
 \begin{align*}
u_\nu({\bf x}, t;{\bf e})&=\mathbb E  \left[f({\bf x}+v {\bf e} Y_\nu(t), {\bf e}) e^{-\alpha Y_\nu(t)}\right]\\
&\quad+\alpha \mathbb E  \left[\int_0^{Y_\nu(t)}e^{-\alpha s}ds \int_{\mathbb S^{d-1}}u_\nu({\bf x}+v{\bf e}s,Y_\nu(t)-s;{\bf e}')\sigma(d{\bf e}')\right]\\
&=\mathbb E  \left[f({\bf x}+v {\bf e} Y_\nu(t), {\bf e}) e^{-\alpha Y_\nu(t)}\right]\\
&\quad+\alpha \int_0^\infty \mu_\nu(u,t)du  \left[\int_0^{u}e^{-\alpha s}\,\mathbb P\, u_\nu({\bf x}+v{\bf e}s,u-s;{\bf e}')ds \right]\\
&=\mathbb E  \left[f({\bf x}+v {\bf e} Y_\nu(t), {\bf e}) e^{-\alpha Y_\nu(t)}\right]\\
&\quad+\alpha \int_0^\infty e^{-\alpha s} ds  \left[\int_s^{\infty}\mathbb P\, u_\nu({\bf x}+v{\bf e}s,u-s;{\bf e}')\mu_\nu(u,t)du \right]\\
&=\mathbb E  \left[f({\bf x}+v {\bf e} Y_\nu(t), {\bf e}) e^{-\alpha Y_\nu(t)}\right]+\mathbb E\left[1_{Z<Y_\nu(t)}\,\mathbb P \,u({\bf x}+v{\bf e}s, Y_\nu(t)-Z;{\bf e}')\right] ,
\end{align*}
where $Z$ is an exponential random variable with rate $\alpha,$ independent of $Y_\nu(t).$

\begin{figure}
        \centering
        \begin{subfigure}[b]{0.475\textwidth}
            \centering
            \begin{overpic}[width=0.8\linewidth]{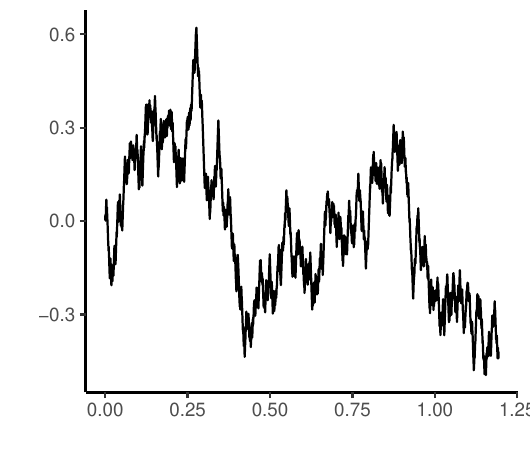}
            \put(-6,67){\scriptsize $X(u)$}
            \put(60,2){\scriptsize $u$}
            \end{overpic}
            \caption[Network2]%
            {{\small ${X}$ ($v=10^{3/2}, \alpha=10^3$).}}    
            \label{fig:mean and std of net14}
        \end{subfigure}
        \begin{subfigure}[b]{0.475\textwidth}  
            \centering 
            \begin{overpic}[width=0.75\linewidth]{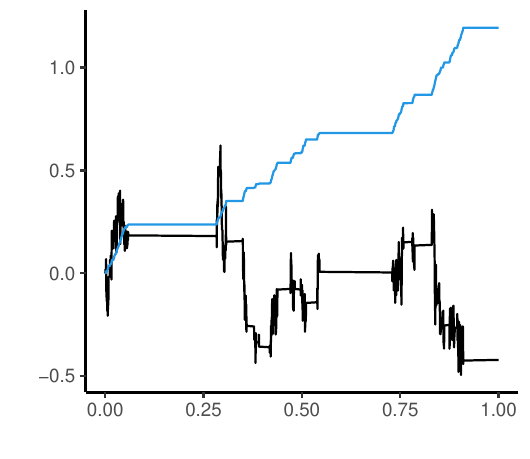}
            \put(80,50){\scriptsize $X(Y_\nu(t))$}
            \put(70, 78){\scriptsize $Y_\nu(t)$}
            \put(55,2){\scriptsize $t$}
            \end{overpic}

            \caption[]%
            {\small ${X}(Y_\nu)$ , $Y_\nu$ ($\nu$=0.75
).}    
            \label{fig:mean and std of net24}
        \end{subfigure}
        \caption[]
        {Simulated paths of  ${ X}$, ${ X}(Y_\nu)$ (black), $Y_\nu$ (blue) for large  $v$ and $\alpha$.} 
        \label{fig:tchange-path-1000}
\end{figure}
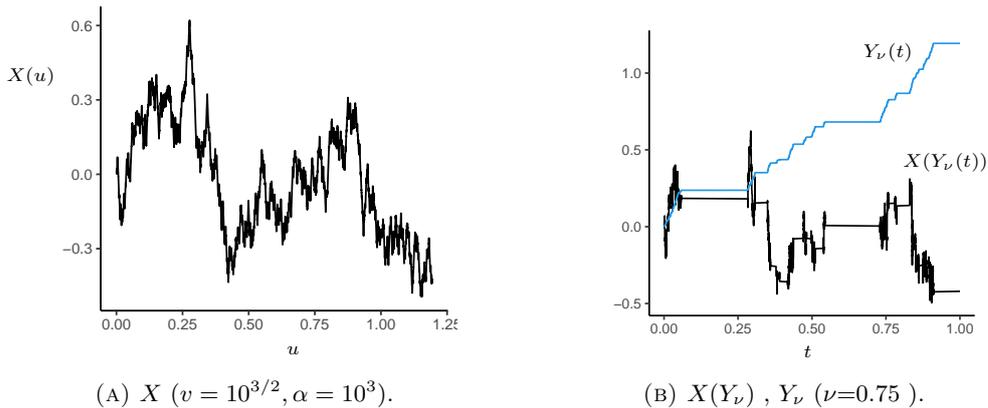

It is not hard to prove that for any $t\geq 0,$ ${\bf X}(Y_\nu(t))$ converges in distribution to ${\bf B}(Y_\nu(t))$ where $\{{\bf B}(t):t\geq 0\}$ is a standard $d$-dimensional Brownian motion.
By assuming that $\frac{\alpha}{v^2}=\frac2d+o(1)$ as $v,\alpha\to\infty,$ we can apply Corollary 
at page 169 in \cite{watanabe1970convergence} and then 
 \begin{align*}
 \lim_{v,\alpha\to\infty}\textsc{P}_{{\bf e}}({\bf x}+{\bf X}(Y_\nu(t))\in A)&= \lim_{v,\alpha\to\infty}\int_0^\infty \textsc{P}_{{\bf e}}({\bf x}+{\bf X}(u)\in A)\mu_\nu(u,t)du\\
 &=\int_0^\infty \textsc{P}({\bf x}+{\bf B}(u)\in A)\mu_\nu(u,t)du\\
 &=\textsc{P}({\bf x}+{\bf B}(Y_\nu(t))\in A) ,
 \end{align*}
where $A\in\mathcal B(\mathbb R^d)$ such that $\partial A$ has Lebesgue measure 0. 
By applying the Portmanteau theorem we can conclude that ${\bf x}+{\bf X}(Y_\nu(t))$ converges weakly to ${\bf x}+{\bf B}(Y_\nu(t))$.
A simulated sample path representing the limiting behaviour of ${X}(Y_\nu)$ is shown in \autoref{fig:tchange-path-1000}, where we set $\alpha=10^3$ and $v = 10^{3/2}$. The simulations were performed as in \autoref{fig:tchange-path}, by setting $\Delta t = 10^{-4}.$ 
As noticed for ${\bf X}(Y_\nu(t))$, the trapping effect of ${\bf B}(Y_\nu(t))$ depends on the values of $\nu$.

\begin{remark}
    It is worth mentioning that alternative anomalous scattering transport processes have been introduced in literature. For instance in \cite{ricciuti2023semi}, the authors deal with a particle switching velocity as in \eqref{eq:vjp}, that is with Mittag-Leffler waiting times, and  having random position given by
    $${\bf X}_\nu(t)={\bf x}+\int_0^t {\bf V}(Y_\nu(s))ds.$$
    Clearly, the previous process has sample paths which differ from those of the random flight \eqref{eq:rptc}, obtained time-changing the position of the particle in the standard case. Indeed, in \cite{ricciuti2023semi} the fractional  Boltzmann equation governing the couple $({\bf X}_\nu(t), {\bf V}(Y_\nu(t)))$ involves a non-local operator which does not coincide with the fractional Caputo derivative. 
    
    In \cite{dicrescenzo2018}, a one-dimensional telegraph process with generalized Mittag-Leffler waiting times has been analyzed.

    Closer to our approach is the random motion studied in \cite{10.1214/EJP.v14-675}, where the authors consider a planar model with time change given by a reflected Brownian motion. 
    Further examples of time-changed random motions related to fractional operators have been discussed in \cite{CINQUE2024128188}.
\end{remark}

\section{Continuous-time random walk} \label{ctrw}
Here we show how it is possible to describe the anomalous run-and-tumble motions in the 
framework of (space-time coupled) continuous-time random walks (CRTW) \cite{klafter2011first,angelani2013averaged}.
The random walk consists of independent steps described by the quantities 
${\mathfrak{\sigma}}({\bf x},t)$, 
the propagator of a completed step 
(space-time probability density to end a step at ${\bf x}$ at time $t$),
and $\Lambda({\bf x},t)$, the propagator of an incomplete step 
(space probability density that  the particle is at ${\bf x}$ 
at time $t$ having not finished the step).
The total pdf to find the particle at ${\bf x}$ at time $t$ can be written as a sum of convolution terms
\begin{equation}
    P_\nu({\bf x},t) = \sum_{n=0}^\infty 
[\underbrace{\sigma * \sigma * \cdots * \sigma}_{n\ \text{times}} * \Lambda]({\bf x},t) ,
\end{equation}
where 
\begin{align*}
[f_1 * \cdots * f_n]({\bf x},t) &= \int_{\mathbb R^{d}\times\cdots\times \mathbb R^{d}} d{\bf x}_1 \cdots d{\bf x}_n\\
&\quad\times\int_0^\infty dt_1 \cdots dt_n \ f_1({\bf x}_1,t_1) \cdots f_n({\bf x}_n,t_n)
\delta\left( \sum_i {\bf x}_i -{\bf x}\right)
\delta\left( \sum_i t_i -t \right) .
\end{align*}
In the Fourier-Laplace domain one has
\begin{equation}
\label{PCTRW}
    {\hat {\tilde P}}_\nu({\bf k},s) = 
    \frac{{\hat {\tilde \Lambda}}({\bf k},s)}{1-{\hat {\tilde \sigma}}({\bf k},s)} .
\end{equation}
The classical run-and-tumble motion is described by a Poisson jump process at constant 
velocity, whose run-time pdf and conditional pdf of displacements given the time $t$ are
\begin{align}
\label{psi}
\psi(t) &= \alpha e^{-\alpha t} ,\\
\label{lambda}
\lambda({\bf x}|t) &= \frac{1}{\Omega_d r^{d-1}} \delta(r-vt) ,
\end{align}
where $r=||{\bf x}||.$
The time-changed procedure allows us to write the propagators of 
the anomalous run-and-tumble motion as
\begin{align}
\label{sig_nu}
    \sigma({\bf x},t) &= \int_0^\infty du \ g_\nu(t,u) \ \lambda({\bf x}|u) \ \psi(u) ,\\
    \Lambda({\bf x},t) &= \int_0^\infty du \ \mu_\nu(u, t)\  \lambda({\bf x}|u)  
    \label{lam_nu}
    \int_u^\infty  \psi(\tau)  d\tau .
\end{align}
We note that for $\nu \to 1$ the functions $g_\nu$ and $\mu_\nu$ tend to a delta function $\delta(t-u)$ and 
the problem reduces to the standard run-and-tumble motion with propagators 
\cite{angelani2013averaged,Detch2014}
\begin{align}
    \sigma({\bf x},t) & \xrightarrow[\nu \to 1]{}
     \lambda({\bf x}|t) \ \psi(t) , \\
    \Lambda({\bf x},t) & \xrightarrow[\nu \to 1]{} 
    \lambda({\bf x}|t)  \int_t^\infty  \psi(u)  du .
\end{align}
Substituting (\ref{psi})-(\ref{lambda}) in the general 
expressions (\ref{sig_nu})-(\ref{lam_nu}) we obtain 
\begin{align}
\label{sig_nu2}
    \sigma({\bf x},t) &= \frac{\alpha}{v \Omega_d r^{d-1}} \ e^{-r \alpha/v} \ g_\nu(t,r/v) ,\\
    \Lambda({\bf x},t) &= \frac{1}{v \Omega_d r^{d-1}} \ e^{-r \alpha/v} \ \mu_\nu(r/v,t) .
\label{lam_nu2}
\end{align}
The underlying random walk process is then characterized by steps with run time distribution
\begin{equation}
    \varphi(t) = \int_{\mathbb R^d} d{\bf x} \ \sigma({\bf x},t) = 
    \int_0^\infty du \ g_\nu(t,u) \ \psi(u) = 
    \alpha t^{\nu-1} E_{\nu,\nu}(-\alpha t^\nu) =
    -\partial_t E_\nu(-\alpha t^\nu)  ,
\label{phit}
\end{equation}
and displacement distribution 
\begin{equation}
\label{PDFrho}
    \rho({\bf x}) = \int_0^\infty dt \ \sigma({\bf x},t) = 
    \int_0^\infty du \ \lambda({\bf x}|u) \ \psi(u) =
    \frac{\alpha}{v \Omega_d r^{d-1}} \ e^{-r \alpha/v}  ,
\end{equation}
not trivially coupled through equations (\ref{sig_nu})-(\ref{lam_nu}).
In deriving (\ref{phit}) we have used the properties
(\ref{L2g}), (\ref{LML})
and $d E_\nu(-x) /dx = - \nu^{-1}E_{\nu,\nu}(-x)$ (see \cite{gorenflo2020mittag}).
We note that the length distribution of the steps (\ref{PDFrho}) is independent of $\nu$
and  is the same of the classical run-and-tumble motion, as the time-change affects only the
run-time of the particle during its motion (becoming a random variable) and not the
length of the space traveled.
We also note that the spatial integral of the quantity $\Lambda({\bf x},t)$ gives the probability 
that the time $T$ between two consecutive jumps exceeds $t$
\begin{equation}
    \textsc{P}(T>t) = \int_{\mathbb R^d} d{\bf x}\ \Lambda({\bf x},t) = 
    \int_0^\infty du \ \mu_\nu(u, t)\ 
    \int_u^\infty  \psi(\tau)  d\tau = {\tilde \mu}(\alpha,t) = E_\nu(-\alpha t^\nu)  ,
\end{equation}
having used (\ref{lam_nu}), (\ref{psi}), (\ref{lambda}) and the property (\ref{L1mu}).\\
By using the known Laplace transforms of $g$ and $\mu$, 
(\ref{L1g}) and (\ref{L1mu}),
we have that the propagators (\ref{sig_nu2})-(\ref{lam_nu2})
in the Fourier-Laplace domains read
\begin{align}
    {\hat {\tilde \sigma}}({\bf k},s) &= \alpha P_0({\bf k},s) , \\
    {\hat {\tilde \Lambda}}({\bf k},s) &= s^{\nu-1} P_0({\bf k},s) ,
\end{align}
where $P_0$ is given by (\ref{P0nu}).
Inserting in (\ref{PCTRW}) we retrieve the solution (\ref{Pks_fr}).

\section{Fractional telegraph equation and its stochastic solution}\label{ftequation}
 We now consider in detail the more interesting case for the applications. The one-dimensional anomalous transport process \eqref{eq:tctp} is directly related to the time-fractional telegraph-type equation. By means of this connection, we are able to provide a clear stochastic interpretation for the solution of the fractional telegraph-type equation. 

For simplicity we set $u_\nu(x,t; {\bf e})=u(x,t; {\bf e})$. For $d=1$, we have that ${\bf e}\in\{-1,1\}$ and the solution \eqref{eq:ssfra} of the Cauchy problem \eqref{fra}, is given by
\begin{align*}
u(x,t; {\bf e})&=\mathbb{E}_{\bf e}[f(x+ X(Y_\nu(t)),V(Y_\nu(t)))] \\
&=\sum_{j\in\{-1, 1\}}\int_{\mathbb R}f(x+y, jv)\int_0^\infty  \textsc{P}( X(u)\in dy, V(u)=jv| V(0)=  {\bf e})\mu_\nu(u,t) du ,
\end{align*}
where $\mu_\nu$ represents the probability density function of $Y_\nu(t)$ at time $t>0.$

The fractional kinetic equation \eqref{fra} leads to the following system involving two time-fractional partial differential equations
\begin{align*}
&\partial_t^\nu u(x,t;1)=v\partial_x  u(x,t; 1)+{ \frac{\alpha}{2}}( u(x,t;-1)- u(x,t; 1)),\\
&\partial_t^\nu u(x,t;-1)=-v\partial_x  u(x,t; -1)+{\frac{\alpha}{2}}( u(x,t;1)- u(x,t; -1)) ,
\end{align*}
and by setting $u(x,t)=u(x,t;1)+u(x,t;-1)$ and $w(x,t)=u(x,t;1)-u(x,t;-1),$ we can write down
\begin{align}\label{eq:fpdes}
&\partial_t^\nu u(x,t)=v\partial_x w(x,t),\\
&\partial_t^\nu w(x,t)=v\partial_x  u(x,t)-{ \alpha} w(x,t) .
\label{eq:fpdes2}
\end{align}

By applying the time-fractional differentiation $\partial_t^\nu$ and the first space derivative to equations \eqref{eq:fpdes}-\eqref{eq:fpdes2}, one has
\begin{align*}
&\partial_t^\nu \partial_t^\nu u(x,t)=v\partial_t^\nu \partial_x w(x,t),\\
&\partial_t^\nu\partial_x  w(x,t)=v\partial_{xx}^2  u(x,t)-{\alpha} \partial_x w(x,t) .
\end{align*}
Therefore, we obtain the fractional telegraph equation 
\begin{align}\label{eq:fte}
&\partial_t^\nu \partial_t^\nu u(x,t)+{ \alpha} \partial_t^\nu u(x,t)=v^2 \partial_{xx}^2  u(x,t),
\end{align}
also studied in \cite{li2019fractional} and \cite{angelani2020fractional}.

We shed in light that $u(x,t)$ represents a solution for \eqref{eq:fte} for any $\nu\in(0,1).$ Furthermore, the equation \eqref{eq:fte} differs from the fractional damped wave equation studied in \cite{orsingher2004time,d2014time}. This is due to the lack of semigroup property for the Caputo derivative and then $\partial_t^\nu \partial_t^\nu\neq \partial_t^{2\nu}.$ Nevertheless, under  suitable initial conditions, it is possible to get the Cauchy problem studied in \cite{orsingher2004time}.

Now, let $\nu\in(\frac12, 1)$ and suppose  $f(x, {\bf e})=\phi(x).$  Hence
the fractional Cauchy problem 
\begin{align}\label{eq:fcp}
&\partial_t^\nu \partial_t^\nu u(x,t)+{ \alpha} \partial_t^\nu u(x,t)=v^2 \partial_{xx}^2  u(x,t) ,\\
& u(x,0)=\phi(x),\quad \partial_t^\nu u(0,x)=0, \notag
\end{align}
is equivalent to the following problem studied, e.g., in \cite{orsingher2004time} or \cite{li2019fractional}
\begin{align}\label{eq:fcp2}
&\partial_t^{2\nu} u(x,t)+{ \alpha} \partial_t^\nu u(x,t)=v^2 \partial_{xx}^2  u(x,t), \\
& u(x,0)=\phi(x),\quad \partial_t u(0,x)=0 ,\notag
\end{align}
(for the proof it is sufficient to apply the properties of Laplace transform for the Caputo derivatives).
Clearly, for $\nu=1$ both time-fractional equations appearing in \eqref{eq:fcp}-\eqref{eq:fcp2}, reduce to the classical telegraph equation \eqref{eq:cpt}. 
We observe that the time fractional telegrapher's
equation \eqref{eq:fcp2} was also derived from the standard telegrapher's
equation by using the subordination approach in \cite{PhysRevE.104.024113} and \cite{gorska2023subordination}.
Moreover, an interesting recent generalization of the telegrapher's equation with power-law memory kernels derived within the persistent random walk theory has been studied in \cite{PhysRevE.102.022128}.

Then, we are able to provide a suitable probabilistic interpretation of the unique solution of the fractional telegraph-type equation in  \eqref{eq:fcp2} (or equivalently in \eqref{eq:fcp}); that is
\begin{align}\label{eq:sste}
u(x,t)&=\mathbb E \left[\phi(x+X(Y_\nu(t)))\right]\\
&=\frac12\left(\mathbb E_{1} \left[\phi(x+ X(Y_\nu(t)))\right]+\mathbb E_{-1} \left[\phi(x+X(Y_\nu(t)))\right]\right)\notag\\
&=\frac12 \left(\mathbb E \left[\phi\left(x+v \int_0^{Y_\nu(t)}(-1)^{N(s) }ds\right)\right]+\mathbb E\left[ \phi\left(x-v \int_0^{Y_\nu(t)}(-1)^{N(s) }ds\right)\right]\right).\notag
\end{align}
Therefore, \eqref{eq:sste} allows to conclude that the anomalous telegraph process is the random model governed by the fractional telegraph equation \eqref{eq:fcp}. Furthermore, \eqref{eq:sste}  generalizes Kac's solution \eqref{eq:kss} time-changing the classical solution.

Going back to the general scheme provided by Eq. \eqref{Pks_fr}, in  dimension $d=1$ we have 
\begin{equation}
P_0(k,s) =\frac{s^\nu+\alpha}{(s^\nu+\alpha)^2 + (vk)^2} ,
\end{equation}
and then
\begin{equation}\label{fsf}
{\hat {\tilde P}}(k,s) =\frac{s^{2\nu-1} +\alpha s^{\nu-1}}{s^{2\nu}+\alpha s^\nu + (vk)^2} .
\end{equation}

We now easily show that the Fourier-Laplace transform of the fundamental solution (\ref{fsf}) coincides with the 
the Fourier-Laplace transform of the Green function for the time-fractional telegraph equation. Indeed, by algebraic manipulation we can write (\ref{fsf}) as follows
\begin{equation}
\left(s^{2\nu}+\alpha s^\nu + (vk)^2\right){\hat {\tilde P}}(k,s) = s^{2\nu-1} +\alpha s^{\nu-1} ,
\end{equation}
and recalling the Laplace transform for Caputo fractional derivatives \eqref{ltfd}, we recognize that the expression in \eqref{fsf} coincides with the Fourier-Laplace transform of the solution for the time-fractional equation
\begin{equation} \label{fc}
\partial_t^{2\nu} u(x,t)+\alpha \partial_t^\nu u(x,t)=v^2 \partial_{xx}^2  u(x,t),
\end{equation}
under the initial conditions $u(x,0) = \delta(x)$ and $\partial_t u(x,0)=0$.
We have two relevant outcomes. First of all, this is the first rigorous proof of the relation between the fractional telegraph equation and the fractional kinetic equation \eqref{eq:fked}. Then, we can say that the stochastic solution of the fractional telegraph process coincides with the time-changed telegraph process \eqref{eq:tctp}.

In the literature it is known the Fourier transform of the solution for \eqref{fc} under the given conditions (see, e.g., \cite{d2014time}) and therefore we can directly obtain the characteristic function of the process ${ \bf X}(Y_\nu(t))$ that coincides for $\nu = 1$ with the characteristic function of the classical telegraph process. 

Moreover, by inverting the Fourier transform, using the property
\begin{equation}
{\mathcal F}^{-1} \Big[\frac{1}{a^2 + k^2}\Big](x) = 
\int_{\mathbb R} \frac{dk}{2\pi} \frac{e^{-ikx}}{a^2+k^2} = 
\frac{1}{2a} \exp(-a|x|) ,
\end{equation}
we obtain
\begin{equation}
{\tilde P}(x,s) = 
\frac{\sqrt{s^{\nu}(s^\nu+\alpha)}}{2vs} \ 
\exp{\Big(-\frac{\sqrt{s^\nu(s^\nu+\alpha)}}{v} |x|\Big)} ,
\end{equation}
in agreement with the result reported in \cite{angelani2020fractional}.
The previous expression can also be obtained from \eqref{eq:ltfdf} using the solution of the 
classical run-and-tumble process.

For simplicity we set $v=1.$ Now, we show that the stochastic solution \eqref{eq:sste} coincides with the representation (4.12) in \cite{li2019fractional} given by
\begin{align}\label{veq:sste2}
   u(x,t)=\mathbb E \phi(x+S_\nu(t)))=\frac12 \left[\mathbb E \phi\left(x+Z_\nu(t)\right)+\mathbb E \phi\left(x- Z_\nu(t)\right)\right]  ,
\end{align}
where $S_\nu(t):=V(0)Z_\nu(t)$ and $\{Z_\nu(t):t\geq 0\}$ represents the inverse of a subordinator with Laplace exponent given by $\sqrt{s^{2\nu}+\alpha s^{\nu}};$ i.e. let $\eta_\nu(z,t), z>0,$ be the density function of $Z_\nu(t),$ we have
$$\int_0^\infty e^{-s t}\eta_\nu(z,t)dt=\frac{\sqrt{s^{2\nu}+{\alpha} s^{\nu}}}{s}e^{-
z\sqrt{s^{2\nu}+{\alpha} s^{\nu}}}.$$
It is not hard to check that the density function of $S_\nu(t), t\geq 0,$ is given by $\frac{1}{2}\eta_\nu(|x|,t), x\in\mathbb R.$ Therefore
$$\frac12\int_0^\infty e^{-s t}\eta_\nu(|x|,t)dt=\tilde P(x,s),$$
and then $S_\nu(t)\stackrel{d}{=}X(Y_\nu(t)).$

We also recall that it is possible to find the
inverse Laplace transform of \eqref{fsf} that is given by 
(see e.g. \cite{orsingher2004time}) 
\begin{equation}
	{\hat P}(k,t) = \frac{1}{2}\bigg[\left(1+\frac{\alpha}{\sqrt{\alpha^2-4 v^2k^2}}\right)E_{\nu}(r_1t^\nu)+\left(1-\frac{\alpha}{\sqrt{\alpha^2-4v^2k^2}}\right)E_{\nu}(r_2 t^\nu)\bigg] ,
\end{equation}
where 
\begin{equation}
	r_1 = -\frac{\alpha}{2}+\sqrt{\frac{\alpha^2}{4}-v^2k^2}, \quad 
 r_2 = -\frac{\alpha}{2}-\sqrt{\frac{\alpha^2}{4}-v^2k^2}.
\end{equation}

\section{Conclusions}

In this paper, anomalous transport models were introduced using a general approach derived from the generalization of the kinetic equation through the time fractional derivative. Nowadays, the fractional operators as well as the fractional partial differential equations represent standard tools to study motions 
different from Brownian motion with the inclusion of memory effects.

In particular, starting from the Kolmogorov forward equation governing the so-called run and tumble walks or random flights,  we have studied its possible generalization  by replacing the time derivative appearing in \eqref{eq:fked} with the Caputo fractional derivative, 
thus introducing memory effects in the kinetic model.
Furthermore, by resorting to the general subordination theory, it is possible to describe the random motions associated to  \eqref{eq:fked} and \eqref{fra} as  time-changed random flights. 
These latter stochastic models are deeply analyzed and many properties highlighted. The time-changed run and tumble walks represent anomalous scattering motions showing a different behaviour with respect to the original random flight: for instance they are no longer finite velocity motions.  
We also provided a description of the anomalous run-and-tumble motion in the framework of continuous-time random walk.
Finally, we have analyzed the one-dimensional case; i.e the fractional telegraph process. It is worth mentioning that we are able to provide a stochastic solution of the fractional telegraph equation, in the true hyperbolic regime $\nu\in(\frac12,1)$, which is given by the original Kac's solution with random clock.  

Some generalizations of the models studied in this paper are possible. Indeed, a future research topic is represented by  anomalous random flights with space-dependent velocity. Inspired by \cite{PhysRevE.100.052147}, we can introduce and analyze a time-fractional version of the telegraph equation with speed  depending on the space $x.$ Another generalization concerns the time-changed run and tumble motions in the stochastic resetting framework, where the particle position is reset randomly in time  to some fixed
point (e.g. its initial position). For an overview with discussion on the applications of the stochastic resetting models the reader can consult, e.g., \cite{Evans_2020}.

\hfill
\section*{Acknowledgments}
LA acknowledges the Italian Ministry of University and Research (MUR) under PRIN2020 Grant No. 2020PFCXPE. The research of ADG is partially supported by  Italian Ministry of University and Research (MUR) under PRIN 2022 (APRIDACAS),
Anomalous Phenomena on Regular and Irregular Domains: Approximating Complexity for the Applied Sciences, Funded by EU - Next Generation EU
CUP B53D23009540006 - Grant Code 2022XZSAFN - PNRR M4.C2.1.1.

\hfill
\section*{Data availability}
All data that support the findings of this study are included within the article.


\appendix 
\section{Time-fractional Cauchy problems and stochastic solutions}
\label{App1}
For the utility of the reader, here we briefly recall the basic mathematical theory about abstract time-fractional Cauchy problems and their stochastic interpretation.
For a complete treatment, we refer for example to \cite{baeumer2001stochastic} and to the recent monograph \cite{ascione2023fractional}.

     First of all, let us recall that a family of linear operators $T_t$, $t\geq 0$ on a 
     Banach space $X$ is called a $C_0$ semigroup if
     \begin{align}
     &T_0 f = f\\
     &T_t T_s f = T_{t+s}f, \\
     &\|T_t f- f\|\rightarrow 0, \quad \mbox{in the Banach space norm as $t\rightarrow 0$}\\
     &\forall t\geq 0, \exists \  \mbox{a constant $M_t>0$ such that} \ \|T_t f\|\leq M_t \|f\|,
     \end{align}
     for all $f\in X$.
     Every $C_0$ semigroup has a generator
     \begin{equation}
     Af = \lim_{t\rightarrow 0}\frac{T_t f- f}{t},
     \end{equation}
     defined for $f \in \mbox{Dom}(A)$.
     
     Then, we recall that $p(x,t) = T_t f(x)$ solves the abstract Cauchy problem 
     \begin{equation}\label{ab1}
     	\partial_t p = A p, \quad p(x, 0) = f(x), \quad \forall f\in \mbox{Dom}(A).
    \end{equation}
    Furthermore, let $\{X(t):t\geq 0\}$ be a Markov process with infinitesimal generator $A,$ we have that the solution of the abstract Cauchy problem \eqref{ab1} is given by
    $$p(x,t)=\mathbb E_x[f(X(t))]$$
    The abstract fractional Cauchy problem involving the Caputo fractional derivative of order $\nu \in (0,1)$ 
    \begin{equation}\label{ab2}
    	\partial_t^{\nu} q = A q, \quad q(x, 0) = f(x), \quad \forall f\in \mbox{Dom}(A),
    \end{equation}
    has solution 
    \begin{equation}
    q(x,t)= \int_0^\infty p(x,u) \mu_\nu(u,t)du,
    \end{equation}
   where $p(x,t)= T_t f(x)$ is the solution of the Cauchy problem \eqref{ab1}, while $\mu_\nu(u,t)$ is the density of the inverse of a stable subordinator $Y_\nu(t)$
   (in the Appendix \ref{App2} we summarize some useful properties of the function
   $\mu_\nu(u,t)$ as well as of the probability density $g_\nu(t,u)$ of the stable
   subordinator $L_\nu(u)$.
By using the property (\ref{muprop1}) we can write the solution in the form
  \begin{equation}
  	q(x,t)= \int_0^\infty p(x,u)\frac{t}{\nu} u^{-1-1/\nu}g_\nu(t u^{-1/\nu},1)du.
  \end{equation} 
  The stochastic representation of the solution of the fractional Cauchy problem
  \eqref{ab2} is the following one
  \begin{equation}\label{eq:ssafcp}
  q(x,t)= \mathbb{E}(p(x,Y_\nu(t))).
  \end{equation}
  One of the most relevant consequence of the general theory is given by 
  the stochastic representation of the solution of the time-fractional heat equation. 
  Let us consider the fractional Cauchy problem 
  \begin{equation}\label{fb}
    	\partial_t^{\nu} u = \frac{1}{2} \partial_{xx} u, \quad u(x, 0) = f(x).
    \end{equation}
    The stochastic representation of the solution is given by
    \begin{equation}
        u(x,t) = \mathbb{E}\bigg[f(x+B(Y_\nu(t))\bigg],
    \end{equation}
    where we denoted by $B(Y_\nu(t))$, the Brownian motion time-changed with the inverse of the stable subordinator $Y_\nu(t)$.
    This means that the fundamental solution of the time-fractional heat equation (that can be represented by means of M-Wright functions, see \cite{mainw}) coincides with the density of the time-changed process $B(Y_\nu(t))$.

\section{Properties of the functions $g_\nu$ and $\mu_\nu$}
\label{App2}
For convenience we summarize here the main properties of the probability density functions
$g_\nu(t,u)$ and $\mu_\nu(u,t)$ of the
stable subordinator $L_\nu(u)$ and the inverse stable subordinator $Y_\nu(t)$
\cite{meerschaert2019inverse}.\\
The function $g_\nu$ satisfies the following scaling relation 
\begin{equation}
g_\nu(t,u)=u^{-1/\nu} g_\nu(tu^{-1/\nu},1)
\end{equation}
The Laplace transform of $g_\nu(t,u)$ with respect to the variables $t$ is
\begin{equation}
\label{L1g}
    {\tilde g}_\nu(s,u) = e^{-us^\nu} .
\end{equation}
The Laplace transform with respect to the variable $u$ is
(we define the Laplace variables pairs
$t \leftrightarrow s$ and $u \leftrightarrow \sigma$)
\begin{equation}
\label{L2g}
    {\tilde g}_\nu(t,\sigma) = t^{\nu-1} \ E_{\nu,\nu}(-\sigma t^\nu) .
\end{equation}
The last expression can be easily obtained by noting that the double Laplace transform of
$g_\nu$ reads -- from (\ref{L1g})
\begin{equation}
    {\tilde {\tilde g}}_\nu(s,\sigma) = \frac{1}{s^\nu+\sigma},
\end{equation}
and considering the inverse-Laplace transform with respect to $s$,
using the property \cite{gorenflo2020mittag}
\begin{equation}
\label{LML}
{\mathcal L}[t^{\nu-1}E_{\mu,\nu}(at^\mu)](s) = s^{\mu-\nu}/(s^\mu-a) ,
        \qquad   {\textrm Re} \mu,\nu>0 .
\end{equation}
The function $\mu_\nu(u,t)$ is given by
\begin{equation}
\label{muprop1}
    \mu_\nu(u,t) = \frac{t}{\nu} u^{-1-1/\nu}g_\nu(tu^{-1/\nu},1) ,
\end{equation}
and  it is related to the function $g_\nu$  through
\begin{equation}
\label{relgmu}
    \nu u \mu_\nu(u,t) = t g_\nu(t,u) .
\end{equation}
The Laplace transform of $\mu_\nu(u,t)$ with respect to the variables $t$ is
\begin{equation}
\label{L1mu}
    {\tilde \mu}_\nu(u,s) = s^{\nu-1} e^{-us^\nu} ,
\end{equation}
as  obtained by using (\ref{relgmu}), (\ref{L1g}) and the property of the Laplace transform 
${\mathcal L}[t f(t)](s) = -\partial_s {\mathcal L}[f(t)](s)$.
The Laplace transform of $\mu_\nu$ with respect to the variable $u$ reads
\begin{equation}
    {\tilde \mu}_\nu(\sigma,t) = E_{\nu}(-\sigma t^\nu) .
\end{equation}
The latter result, as before for the $g$ function, can be obtained by 
noting that the double Laplace transform of $\mu_\nu$ reads -- see (\ref{L1mu})
\begin{equation}
    {\tilde {\tilde \mu}}_\nu(\sigma,s) = \frac{s^{\nu-1}}{s^\nu+\sigma},
\end{equation}
and considering the inverse-Laplace transform with respect to $s$,
using (\ref{LML}) and the identity $E_{\nu,1}(x)=E_\nu(x)$.\\
Some interesting asymptotic behaviors of the  $\mu_\nu$ function are
as follows.
For fixed $u>0$ and $t\downarrow 0$ we have 
 \begin{equation}
\mu_\nu(u,t)\sim \sqrt{\frac{\nu}{2\pi(1-\nu)}}\left(\frac{\nu}{u}\right)^{\frac{2-\nu}{2-2\nu}}t^{-\frac{\nu}{2-2\nu}}\exp\left(-|1-\nu|u^{\frac{1}{1-\nu}}\left(\frac{\nu}{t}\right)^{\frac{\nu}{1-\nu}}\right),
 \end{equation}
For $t\rightarrow +\infty$, we have that
 \begin{equation}
\mu_\nu(u,t)\sim \frac{t^{-\nu}}{\Gamma(1-\nu)}.
 \end{equation}

 \noindent
 Finally we note that for $\nu\to 1$ the $g_\nu$ and $\mu_\nu$ functions tend to a 
 delta function
 \begin{equation}
     \lim_{\nu \to 1} g_\nu(t,u) = \lim_{\nu \to 1} \mu_\nu(u,t) = \delta(t-u) ,
 \end{equation}
as simply obtained by considering the inverse Laplace transform of 
(\ref{L1g}) and (\ref{L1mu}) for $\nu=1$.


 	\bibliographystyle{abbrv}

\end{document}